\documentclass[aps, prab, reprint, superscriptaddress, nofootinbib, english, floatfix]{revtex4-2}

\usepackage{graphicx}
\usepackage{dcolumn}
\usepackage{amsmath}
\usepackage{amssymb}
\usepackage{mathtools}
\usepackage{xfrac}
\usepackage{hyperref}
\usepackage[binary-units]{siunitx}
\usepackage[acronym]{glossaries}
\usepackage{physics}
\usepackage{csquotes}
\usepackage[caption=false]{subfig}

\providecommand{\rf}{\mathrm{rf}}

\providecommand{\hc}{\mathrm{HC}}
\providecommand{\eff}{\mathrm{eff}}
\providecommand{\mymax}{\mathrm{max}}
\newacronym[plural=MCs,firstplural=main cavities (MCs), longplural={main cavities}]{MC}{MC}{main cavity}
\newglossaryentry{ptbl}{
    name={PTBL},
    description={periodic transient beam loading},
    first={periodic transient beam loading (\glsentrytext{ptbl})}
}
\newglossaryentry{hc}
{
    name={HC},
    description={harmonic cavity},
    first={harmonic cavity (\glsentrytext{hc})},
    plural={HCs},
    descriptionplural={harmonic cavities},
    firstplural={harmonic cavities (\glsentryplural{hc})}
}
\newglossaryentry{lmci}{
    name={LMCI},
    description={longitudinal mode coupling instability},
    first={longitudinal mode coupling instability (\glsentrytext{lmci})}
}

\begin{document}
\title{Theoretical models for longitudinal coupled-bunch instabilities driven by \\ harmonic cavities in electron storage rings}

\author{Murilo B. Alves}
\email{murilo.alves@lnls.br}
\affiliation{Brazilian Synchrotron Light Laboratory -- LNLS, Brazilian Center for Research in Energy and Materials -- CNPEM, 13083-970, Campinas, SP, Brazil.}
\affiliation{Gleb Wataghin Institute of Physics, University of Campinas -- UNICAMP, 13083-859, Campinas, SP, Brazil}
\date{\today}

\begin{abstract}
    We present a theoretical framework for analyzing longitudinal coupled-bunch instabilities in double-rf systems with even filling patterns, accounting for potential-well distortion and multiple azimuthal modes. The linearized Vlasov equation is solved in the frequency-domain for an arbitrary rf potential to derive the Lebedev equation. We unified different formulations, obtaining results from recent publications as particular cases. Applications to Robinson dipole-quadrupole mode coupling and the periodic transient beam loading (PTBL)/mode-1 instability are presented. Notably, for the first time, theoretical predictions of the mode-1 thresholds show excellent agreement with experimental data. The analysis reveals that the PTBL instability is a zero-frequency effect dependent on azimuthal mode interactions and resistant to Landau damping, providing new insights into its mechanism. The methods are implemented in the open-source package \texttt{pycolleff}, offering a useful semi-analytical tool for studying instabilities in electron storage rings with harmonic cavities.
\end{abstract}

\maketitle

\section{Introduction}
Collective instabilities in double-rf systems with \glspl{hc} have been a concern for synchrotrons for many years~\cite{Chin1983, Chin1982, Krinsky1985, Shaposhnikova1994}. In modern synchrotron light sources, passive \glspl{hc} are employed in the bunch lengthening mode to reduce the bunch charge density, alleviating collective effects such as intrabeam scattering, Touschek scattering and impedance-induced heating of components. These effects have become critical issues for the 4th-generation synchrotron light sources with ultralow transverse emittances~\cite{Leemann2014, Nagaoka2014}.

The theoretical description of instabilities in double-rf systems is significantly more challenging than the theory for single-rf systems, for which a well-developed theory exists in the literature~\cite{Sacherer1972, Chao1993, Ng2006}. In electron rings with single-rf systems, assuming linear single-particle dynamics (harmonic rf potential) and Gaussian bunch distributions are valid approximations for short-bunches. These approximations considerably simplify the analytical description and calculations of instabilities. In contrast, the single-particle motion in double-rf systems can be significantly modified by potential-well distortion effects induced by the \glspl{hc}, leading to highly nonlinear dynamics and non-Gaussian bunch distributions~\cite{Hofmann1980,Bassi2019,Venturini2018,Venturini2018b,Lindberg2018}. The effects of \glspl{hc} on beam stability can be twofold: they can stabilize the beam by lengthening the bunches, reducing charge density and providing Landau damping through the spread of incoherent synchrotron frequencies; or they can degrade the stability by lowering the average synchrotron frequency and by introducing additional impedance.

For longitudinal coupled-bunch instabilities in single-rf systems, a standard approach is to ignore the interaction between azimuthal (synchrotron) modes. This simplification assumes that the current per bunch is not too high and that multibunch instabilities are typically driven by narrowband impedances (long-range wakefields) that do not significantly affect intrabunch motion. Under these conditions, the azimuthal modes are sufficiently separated, allowing each mode to be studied independently. However, for coupled-bunch instabilities in double-rf systems, the situation may change. Even for low currents per bunch, the azimuthal modes may interact due to the flattening of the rf potential. Therefore, taking into account the potential-well distortion, which was normally required only to study single-bunch instabilities~\cite{Oide1990,Oide1995,Chao1996,Dyachkov1995,Ng1995,Mosnier1999,Cai2011,Karpov2023}, might also be important~\cite{Venturini2018, Lindberg2018}.

This work develops a theoretical formulation in the frequency-domain to analyze longitudinal instabilities in double-rf systems, accounting for the nonlinear effects of potential-well distortion and interactions between multiple azimuthal modes. The theory applies to both multibunch and single-bunch cases, assuming an even filling pattern such that every filled bucket sees the same equilibrium potential and has the same bunch distribution. Even though the cases discussed in this work are focused on double-rf systems, the framework is also suited for instability studies involving generic narrowband resonators, such as HOMs from rf cavities, while incorporating the impact of potential-well distortion from the machine broadband impedance.

The paper is structured as follows. Sec.~\ref{sec:theory} develops the theoretical models, starting with the linearized Vlasov equation to derive the Sacherer integral equation for arbitrary rf potentials, extending Venturini's results~\cite{Venturini2018}, and demonstrating their equivalence to the Lebedev equation~\cite{Lebedev1968, Karpov2021}. We introduce an effective model and a Gaussian \gls{lmci} theory~\cite{Suzuki1983} adapted for double-rf systems. In Sec.~\ref{sec:dispersion-relation}, a generic dispersion-relation for narrowband resonators is derived, yielding the models from Refs.~\cite{Lindberg2018, Cullinan2022} as particular cases. Sec.~\ref{sec:applications} applies the theory to Robinson dipole-quadrupole mode coupling and \gls{ptbl}/mode-1 instabilities, benchmarking the predictions with MAX IV experimental data and achieving, for the first time, excellent agreement with measured mode-1 thresholds. In Sec.~\ref{sec:discussion}, details of the \gls{ptbl} instability mechanism are discussed. Section~\ref{sec:conclusion} summarizes the findings and presents the conclusions.
\section{Theory\label{sec:theory}}
We will adopt the definition of the longitudinal coordinate~$z$ of relativistic particles in a storage ring with $z>0$ for trailing particles. All the following derivations assume an even filling condition, i.e., all filled buckets with the same current and identical equilibrium longitudinal bunch distributions~$\lambda_0(z)$.

Consider that the longitudinal equilibrium is obtained as a self-consistent solution of Ha{\"{i}ssinski equation considering potential-well distortion effects, for example with the semi-analyical method presented in Ref.~\cite{Alves2023}. This calculation provides the equilibrium wake voltage $V_\mathrm{wake}(z; \lambda_0)$ that is added to the external rf voltage $V_\rf(z)$ to result in the total equilibrium voltage $V_0(z) = V_\rf(z) + V_\mathrm{wake}(z; \lambda_0)$. The equilibrium potential is then calculated as:
\begin{equation}
\Phi_0(z; \lambda_0) = -\frac{1}{E_0C_0}\int_{0}^z {\dd{z^{\prime}} \left[eV_0(z^{\prime}; \lambda_0) - U_0\right]},
\end{equation}
with~$\lambda_0(z)$ satisfying the Ha{\"{i}ssinski equation:
\begin{equation}
    \lambda_0(z) = \frac{1}{\mathcal{N}_z} \exp\left[-\frac{\Phi_0(z; \lambda_0)}{\alpha \sigma_\delta^2}\right],
    \label{eq:zdistribution}
\end{equation}
where the constant~$\mathcal{N}_z$ normalizes $\lambda_0(z)$ to unity, $\sigma_\delta$ is the equilibrium relative energy spread, $\alpha$ is the momentum compaction factor (assuming above transition so the slip factor is $\alpha - 1/\gamma^2 \approx \alpha$), $E_0$ is the ring nominal energy, $C_0$ the ring circumference, $e>0$ is the elementary charge and $U_0$ is the energy loss per turn from synchrotron radiation.

Considering $(z, \delta)$ as canonical coordinates, where $\delta~=~(E-E_0)/E_0$ is the relative energy deviation, in this equilibrium potential the single-particle equations of motion are:
\begin{align}
    \frac{\dd z}{\dd s} &= \alpha \delta, \\
    \frac{\dd \delta}{\dd s} &= \frac{eV_0(z; \lambda_0)-U_0}{E_0C_0},
\end{align}
which are associated with the unperturbed Hamiltonian
\begin{equation}
    \mathcal{H}_0 =  \frac{\alpha \delta^2}{2} + \Phi_0(z; \lambda_0).
    \label{eq:hamiltonian}
\end{equation}

It is useful for the following instability analysis to perform a canonical transformation to action-angle variables~$(z, \delta) \to (J, \varphi)$. The numerical determination of the canonical transformation can be done, for instance, following the procedure described in the Appendix C of Ref.~\cite{Venturini2018}. With that procedure we obtain the transformation in a rectangular grid $z_{ij} = \zeta(J_i, \varphi_j)$.

The two-dimensional distribution $\Psi$ in the longitudinal phase-space satisfies the Vlasov equation:
\begin{equation}
\dv{\Psi}{s} = \pdv{\Psi}{s} + \left\{\Psi, \mathcal{H}\right\} = 0,
\end{equation}
with $\left\{\Psi, \mathcal{H}\right\}$ denoting the Poisson brackets.

We consider a small perturbation from the equilibrium that oscillates with a complex coherent frequency $\Omega$. The perturbed distribution will be associated to a perturbation in the Hamiltonian:
\begin{align}
    \Psi(J, \varphi, s) &=  \Psi_0(J) + \Psi_1(J, \varphi)e^{-i\Omega s/c}, \\
    \mathcal{H}(J, \varphi, s) &= \mathcal{H}_0(J) + \Phi_1(J, \varphi)e^{-i\Omega s/c}\label{eq:hamiltonian-perturb},
\end{align}
where $\Psi_0(J) = (2\pi \mathcal{N}_J)^{-1}e^{-\mathcal{H}_0(J)\slash \alpha \sigma_\delta^2}$ is the equilibrium distribution, also normalized to unity.

Applying this perturbation to the Vlasov equation and linearizing it, reads to
\begin{equation}
    - i\Omega \Psi_1 + \omega_s(J)\pdv{\Psi_1}{\varphi} - c\pdv{\Psi_0}{J}\pdv{\Phi_1}{\varphi} = 0,
\end{equation}
where $\omega_s(J) = c \pdv{\mathcal{H}_0}{J}$ is the amplitude-dependent synchrotron frequency.

For the even filling case, in the equilibrium state all bunch profiles are identical. However, when a coupled-bunch instability is driven, each bunch can have a different profile and time evolution, governed by a system of coupled Vlasov equations. We will assume there are $M$ equidistant bunches in the ring, with $1 \leq M \leq h$, where $h$ is the harmonic number. The perturbation distribution for the $n$th bunch is represented as~$\Psi_{1, n}^{(\ell)}(J, \varphi, s) = \Psi_{1}^{(\ell)}(J, \varphi, s)e^{2\pi i n \ell /M}$ with $\ell = 0, 1, \ldots, M-1$ referring to the coupled-bunch mode number.  Using the coupled-bunch mode basis, $\left\{\ell \right\}$, instead of the bunch index basis, $\left\{n\right\}$, decouples the system of Vlasov equations into $M$ independent equations for each coupled-bunch distribution $\Psi_{1}^{(\ell)}(J, \varphi, s)$. For brevity,  we will drop the reference to the coupled-bunch index $\ell$ in the perturbation distribution. For the single-bunch case, $\ell = 0$ and there is only one Vlasov equation to be solved.

The wake voltage induced by the perturbation is
\begin{equation}
    V_1(z; \lambda_1) = -I_0\sum_{p=-\infty}^{\infty}\tilde{\lambda}_{1; {p,\ell}}(\Omega)Z_{p, \ell}(\Omega)e^{-i\omega_{p,\ell}z/c},
    \label{eq:induced-voltage}
\end{equation}
where $I_0$ is the total beam current, $\omega_{p, \ell} = (pM + \ell)\omega_0$, $\omega_0$ the revolution frequency. For derivations of Eq.~\eqref{eq:induced-voltage}, see Refs.~\cite{Karpov2021, Alves2023}. For compactness, we introduced the notation $\tilde{\lambda}_{1; {p,\ell}}(\Omega):= \tilde{\lambda}_{1}(\omega_{p,\ell} + \Omega)$ and $Z_{p, \ell}(\Omega):=Z(\omega_{p,\ell} + \Omega)$. $\tilde{\lambda}_1(\omega)$ is the Fourier transform of the bunch distribution
\begin{align}
    \tilde{\lambda}_1(\omega) &= \int_{-\infty}^\infty \dd{z} e^{i\omega z/c} \lambda_1(z) \nonumber \\
    &= \int_{-\infty}^\infty \dd{z} e^{i\omega z/c} \int_{-\infty}^\infty \dd{\delta} \Psi_1(z, \delta) \nonumber \\
    &= \int_{0}^\infty\int_{0}^{2\pi} \dd{\varphi} \dd{J} e^{i\omega \zeta(J, \varphi)/c} \Psi_1(J, \varphi).
    \label{eq:bunch_spectrum}
\end{align}

The approximation $\tilde{\lambda}_{1; {p,\ell}}(\Omega) \approx \tilde{\lambda}_{1}(\omega_{p,\ell})$ can generally be done in Eq.~\eqref{eq:induced-voltage}, because $\tilde{\lambda}_{1}(\omega)$ is a smooth function and $\Re(\Omega) \ll \omega_{p,\ell}$. As the impedance $Z(\omega)$ can be related to narrowband resonators, it is important to keep the $\Omega$ dependence in its argument. Note that the term $e^{-i\Omega s/c}$ has already been factored out in Eq.~\eqref{eq:hamiltonian-perturb}. $Z(\Omega)$ is well-defined for complex $\Omega$, given that the impedance function is analytic~\cite{Ng2006}. The corresponding perturbation of the wake potential and its derivative are:
\begin{align}
    \Phi_1(\zeta) &= -\int_{0}^{\zeta}\dd{\zeta^{\prime}}\frac{eV_1(\zeta^{\prime}; \lambda_1)}{E_0C_0}, \\
    \pdv{\Phi_1}{\varphi} &= \frac{eI_0}{E_0C_0}\sum_{p=-\infty}^{\infty} \tilde{\lambda}_1(\omega_{p, \ell})\frac{Z_{p, \ell}(\Omega)}{-i\omega_{p, \ell}/c}\pdv{}{\varphi}e^{-i\omega_{p, \ell} \zeta /c}.
\end{align}

Next, we use the azimuthal symmetry with respect to $\varphi$ to expand the perturbation in azimuthal modes $m$:
\begin{equation}
    \Psi_1(J,\varphi) = \sum_{m\neq 0} R_m(J) e^{im\varphi},
\end{equation}
where $R_m(J)$ are real-valued functions. The bunch spectrum from Eq.~\eqref{eq:bunch_spectrum} is then written as
\begin{align}
    \tilde{\lambda}_1(\omega_{p}) &= 2\pi \sum_{m \neq 0} \int_{0}^\infty \dd{J} R_m(J) H_{m, p}(J),\\
    H_{m, p}(J) & := \frac{1}{2\pi} \int_{0}^{2\pi} \dd{\varphi} e^{im\varphi + i\omega_{p, \ell} \zeta(J, \varphi)/c}.\label{eq:hmp}
\end{align}
As remarked in Ref.~\cite{Karpov2021}, the functions $H_{m, p}(J)$ were first introduced by Lebedev in 1968~\cite{Lebedev1968}. The functions $H_{m, p}(J)$ depend on the beam current and impedances as the canonical transformation $\zeta(J, \varphi)$ is modified by the potential-well distortion.

Inserting all these results in the linearized Vlasov equation, multiplying by $e^{-in\varphi}$ and integrating over $\varphi$ (recall that $\int_{0}^{2\pi}\dd{\varphi} e^{i(m-n)\varphi} = 2\pi \delta_{mn}$) results in
\begin{align}
 & (\Omega - m \omega_s(J) ) R_m(J) + im\kappa \pdv{\Psi_0}{J}\sum_{p=-\infty}^ {\infty} \frac{Z_{p, \ell}(\Omega)}{\omega_{p, \ell}} H_{m, p}^\ast (J)\nonumber \\
   &\times \sum_{m^{\prime}\neq 0} \int_{0}^\infty \dd{J} R_{m^{\prime}}(J) H_{m^{\prime}, p}(J) = 0,
\label{eq:full_vlasov}
\end{align}
where we defined the intensity parameter:
\begin{equation}
    \kappa = \frac{2\pi e I_0 c^2}{E_0 C_0},
    \label{eq:kappa}
\end{equation}
and used the result
\begin{equation}
    \frac{1}{2\pi}\int_{0}^{2\pi}\dd{\varphi} e^{-im\varphi}\pdv{}{\varphi} e^{-i\omega_{p, \ell} \zeta /c} = im H_{m, p}^\ast (J).
\end{equation}

Until this point, our derivation closely followed the notation and steps presented in Venturini's paper, e.g., compare Eq.~\eqref{eq:full_vlasov} here with Eq.~(17) in~\cite{Venturini2018}. The goal now is to further manipulate the integral equation~\eqref{eq:full_vlasov} to obtain a dispersion-relation that must be solved for the coherent frequency $\Omega$.

We will first, for completeness, reproduce Venturini's approach. Then, we will present an extension of Venturini's results that leads to the Lebedev equation. From that, we will introduce an effective model that neglects the Landau damping and present the Gaussian \gls{lmci} model discussed in previous investigations~\cite{Alves2024HarmonLIP,Alves2024IPAC}.
\subsection{Venturini's approach}
Multiplying Eq.~\eqref{eq:full_vlasov} by $H_{m, p^{\prime}}(J)$, dividing it by $\left(\Omega - m\omega_s(J)\right)$ and integrating it over $J$, we obtain:
\begin{equation}
X_{m, p^{\prime}} + im\kappa \sum_{p=-\infty}^{\infty} \frac{Z_{p, \ell}(\Omega)}{\omega_{p, \ell}}G_{m, pp^{\prime}}(\Omega)\sum_{m^{\prime}\neq 0}X_{m^{\prime}, p} = 0,
\label{eq:system_mpp}
\end{equation}
where
\begin{align}
     X_{m, p} &= \int_{0}^\infty \dd{J} R_{m}(J) H_{m, p}(J), \label{eq:xmp} \\
     G_{m, pp^{\prime}} &= \int_{0}^{\infty} \dd{J} \pdv{\Psi_0}{J} \frac{H_{m, p^{\prime}} (J) H_{m, p}^\ast (J)}{\Omega - m\omega_s(J)}.
\end{align}

Equation~\eqref{eq:system_mpp} can be rewritten as an infinite system of equations:
\begin{align}
\sum_{p=-\infty}^{\infty}\sum_{m^{\prime} \neq 0}B_{mm^{\prime}, pp^{\prime}}(\Omega) X_{m^{\prime}, p} = 0, \\
B_{mm^{\prime}, pp^{\prime}}(\Omega) = \delta_{mm^{\prime}, pp^{\prime}} + im \kappa \frac{Z_{p, \ell}(\Omega)}{\omega_{p, \ell}} G_{m, pp^{\prime}}(\Omega).
\end{align}
As in the case of interest of Ref.~\cite{Venturini2018}, for a narrowband resonator, only a single harmonic $\pm p_0$ has a significant impedance contribution. Moreover, in practice, the problem is solved by truncating the sum to $\pm m_{\mymax}$. In this way, $B_{mm^{\prime}, pp^{\prime}}(\Omega)$ becomes a finite $4m_{\mymax} \times 4m_{\mymax}$ matrix. As we are interested in non-trivial solutions, $X_{m, p} \neq 0$, the coherent frequency $\Omega$ is computed as the root of the determinant of the $B(\Omega)$ matrix.

\subsection{Lebedev equation}
Equation~\eqref{eq:system_mpp} can be further simplified. Applying a summation over $m$ and defining
\begin{equation}
    Y_p = \sum_{m\neq 0} X_{m, p} \qq{and} G_{pp^{\prime}}(\Omega) = \sum_{m\neq 0}mG_{m, pp^{\prime}}(\Omega),
\end{equation}
simplifies Eq.~\eqref{eq:system_mpp} to
\begin{equation}
    Y_{p^{\prime}} +  i\kappa \sum_{p=-\infty}^{\infty} \frac{Z_{p, \ell}(\Omega)}{\omega_{p, \ell}}G_{pp^{\prime}}(\Omega)Y_{p} = 0.
\end{equation}

The infinite system of equations is now
\begin{align}
    \sum_{p=-\infty}^{\infty}B_{pp^{\prime}}(\Omega) Y_{p} = 0, \\
    B_{pp^{\prime}}(\Omega) = \delta_{pp^{\prime}} + i\kappa \frac{Z_{p, \ell}(\Omega)}{\omega_{p, \ell}} G_{pp^{\prime}}(\Omega).
\end{align}
In this format, the system of equations is equivalent to the Ledebev equation~\cite{Lebedev1968} as derived in Ref.~\cite{Karpov2021}, see Eqs.~(33-36) from~\cite{Karpov2021}.

From Eq.~\eqref{eq:hmp} we can derive the property $H_{-m, p}(J) = H_{m, p}(J)$, using that $\zeta(J, \varphi)$ is an even function and $2\pi$-periodic with respect to $\varphi$. This symmetry simplifies the matrix $G_{pp^{\prime}}(\Omega)$, combining positive and negative azimuthal modes:
\begin{align}
    G_{pp^{\prime}}(\Omega) &= \int_{0}^{\infty} \dd{J} \pdv{\Psi_0}{J}g_{pp^{\prime}}(J, \Omega), \label{eq:gpp_mpos} \\
    g_{pp^{\prime}}(J, \Omega) &= \sum_{m=1}^{\infty} 2m^2\omega_s(J)\frac{H_{m, p^{\prime}} (J) H_{m, p}^\ast (J)}{\Omega^2 - m^2\omega_s^2(J)},
\end{align}
where it was assumed that the integrand does not diverge so the sum over $m$ can be interchanged with the integral.

We introduced the auxiliary function $g_{pp^{\prime}}(J, \Omega)$. Equation~\eqref{eq:gpp_mpos} for $G_{pp^{\prime}}(\Omega)$ is quite convenient since the truncation of azimuthal modes can be controlled based on the convergence of the function $g_{pp^{\prime}}(J, \Omega)$ at run-time for each iteration of the root finding algorithm for $\Omega$.

It is important to highlight that with the Lebedev equation, the dimensionality of the matrix and the number of numerical integrations do not depend on the truncation $m_{\mymax}$. Hence, we showed that Venturini's formulation is essentially equivalent to the Lebedev equation with the disadvantage of having an avoidable computational complexity that increases with $m_{\mymax}$.

\subsection{Effective synchrotron frequency model\label{subsec:effective-model}}
What set the requirement of a nonlinear solution method for $\Omega$ in the integral equation in Eq.~\eqref{eq:full_vlasov} are the dependencies of the synchrotron frequency with action $\omega_s(J)$ and the impedance with $\Omega$. With this observation, we will formulate a simplified linear problem with minimal changes.

Suppose that $\omega_s(J)$ is replaced by a constant effective synchrotron frequency $\bar{\omega}_s$. A possible choice for $\bar{\omega}_s$ will be presented in the next section. This change may impact the results by neglecting the frequency spread and Landau damping. Additionally, if we approximate $Z_{p, \ell}(\Omega) \approx Z_{{p, \ell}}\left(m\bar{\omega}_s\right)$, then Eq.~\eqref{eq:full_vlasov} can be simplified to an eigenvalue equation:
\begin{align}
    & \sum_{p=-\infty}^{\infty}\sum_{m^{\prime}=-\infty}^{\infty}C_{mm^{\prime}, pp^{\prime}} X_{m^{\prime}, p} = \Omega X_{m, p^{\prime}}, \\
    C_{mm^{\prime}, pp^{\prime}} &=  m \bar{\omega}_s \delta_{mm^{\prime}, pp^{\prime}} - i m \kappa \frac{Z_{{p, \ell}}\left(m\bar{\omega}_s\right)}{\omega_{p, \ell}} F_{m, pp^{\prime}}, \\
    F_{m, pp^{\prime}} &= \int_{0}^{\infty} \dd{J} \pdv{\Psi_0}{J} H_{m, p^{\prime}} (J) H_{m, p}^\ast (J),
\end{align}
and $X_{m,p}$ is the same as defined in Eq.~\eqref{eq:xmp}. We can proceed by truncating the azimuthals $m_\mymax$ and selecting the harmonics $p$ to find the coherent frequencies $\Omega$ as the eigenvalues of $C_{mm^{\prime}, pp^{\prime}}$.

Note that the nonlinearities can still play a role in this model, not through Landau damping as the frequency spread is neglected, but through the terms $H_{m,p}(J)$ that encodes the nonlinear dynamics in the phase of $e^{i\omega_{p, \ell} \zeta(J, \varphi)/c}$~\cite{Schenk2018}. Besides, the bunch profile distortions are accounted through the numerical solution of the Ha{\"{i}ssinski equation, which is used to compute $\pdv{\Psi_0}{J}$ numerically instead of an analytical distribution, as done in the next model we will present.
\subsection{Gaussian longitudinal mode coupling\label{subsec:gaussian-lmci}}
Multibunch instability thresholds can be computed by employing Suzuki's frequency-domain solution of Vlasov equation for longitudinal instabilities, which allows mode coupling between different azimuthal and radial modes of the bunch motion~\cite{Suzuki1983}. The theory assumes that the single-particle dynamics is linear and that the longitudinal bunch distribution is Gaussian. This makes the theory suitable for studying instabilities in single-rf systems, neglecting potential-well distortion. Nevertheless, in the past some success was achieved in using a linear Gaussian theory to study the instabilities in \gls{hc} systems~\cite{Bosch2001, Bassi2019IPAC}.

Suzuki expanded the radial function $R(J)$ in a basis of orthogonal functions to solve Sacherer's integral equation. With Gaussian bunch distributions, generalized Laguerre polynomials were used as orthogonal functions. Suzuki's solution yields the infinite matrix equation~\cite{Suzuki1983}:
\begin{align}
& \sum_{m^\prime=1}^{\infty} \sum_{k^\prime = 0}^{\infty} A_{m^\prime k^\prime}^{m k} b_{k^\prime}^{(m^\prime)} = \left(\frac{\Omega}{\omega_s}\right)^2 b_{k}^{(m)}, \\
    A_{m^\prime k^\prime}^{m k} &= m^2 \delta_{m^\prime m}\delta_{k^\prime k} + i \frac{m^2 ec^2 \alpha I_0}{\pi \sigma_z^2 \omega_s^2 E_0}  M_{m^\prime k^\prime}^{m k},
\end{align}
where $(m, m^\prime)$ and $(k, k^\prime)$ are indices for the azimuthal and radial modes, respectively. The coupling matrix depends on the longitudinal impedance~$Z(\omega)$ and beam spectrum
\begin{align}
   M_{m^\prime k^\prime}^{m k} &= \sum_{p=-\infty}^{\infty} \frac{Z(\omega_{p, \ell}+\Omega)}{\omega_{p, \ell}} i^{m^\prime-m} \nonumber \\
   &~\times I_{m^\prime k^\prime}\left(\frac{\omega_{p, \ell}+\Omega}{\omega_0}\right) I_{m k}\left(\frac{\omega_{p, \ell}+\Omega}{\omega_0}\right).
\end{align}
For Gaussian bunches, the functions $I_{m k}(p)$ have the analytic form:
\begin{equation}
    I_{m k}\left(\frac{\omega_{p, \ell}}{\omega_0}\right)= \frac{1}{\sqrt{(m+k)!k!}} \left(\frac{\zeta_{p, \ell}}{2}\right)^{m+2k} \exp\left({-\frac{\zeta_{p, \ell}^2}{4}}\right),
\end{equation}
where $\zeta_{p, \ell} = \sqrt{2}\sigma_z \omega_{p, \ell}/c$. To solve the matrix problem, the sums are truncated to $m_\mymax$ and $k_\mymax$. Moreover, the approximation $\Omega \approx m\omega_s$\footnote{The approximation $\Omega \approx m\omega_s$ with $m=1$ for all elements was considered in the \texttt{pycolleff} implementation for computational speed, and it was verified that varying $m$ from $0$ to $10$ did not affect the results presented in this paper.} is applied to compute the finite coupling matrix $M_{m^\prime k^\prime}^{m k}$. The analysis can be specialized to each coupled bunch mode $\ell$. Then, the coherent frequencies $\Omega$ are obtained by diagonalization.

The \gls{lmci} theory can be applied to coupled-bunch instabilities in double-rf system, requiring a minor yet important adaptation in the calculation process. The values for bunch length and average incoherent synchrotron frequency can be obtained from the self-consistent solution of the Ha{\"{i}ssinski equation. With this adaptation, the potential-well distortion caused by the \gls{hc} is not fully neglected for the instability analysis. However, it is important to note that this scheme also ignores the frequency spread, thus Landau damping effects are neglected. We will refer to this approximate model as \enquote{Gaussian \gls{lmci}}.

Such as in the effective synchrotron frequency model, in the Gaussian \gls{lmci} the constant incoherent synchrotron frequency is a crucial input. Considering that the approximation of Gaussian bunch is already made, a simple choice for the constant frequency is to maintain the relation between synchrotron frequency and bunch length that holds for harmonic single-rf systems (quadratic rf potential):
\begin{align}
    \langle\omega_s\rangle_\mathrm{quadratic} &= \frac{\alpha c \sigma_\delta}{\sigma_z}\label{eq:freq_by_bunlen}.
\end{align}

The synchrotron frequency can be determined by the bunch length (assuming the momentum compaction and energy spread are fixed). In this way, we will be evaluating the instability in a fictitious equivalent quadratic system with the same bunch length as produced by the \gls{hc}. Such approach was suggested in Refs.~\cite{Lindberg2016,Lindberg2018}.

The Gaussian \gls{lmci} method has the advantage of being considerably faster than the previous methods of solving the Lebedev equation and the effective synchrotron frequency model, since its matrix elements are computed by analytical expressions, while the others require additional calculations for the numerical canonical transformation and numerical integrations. Equation~\eqref{eq:freq_by_bunlen} will also be the choice for $\bar{\omega}_s$ in the effective model from Sec.~\ref{subsec:effective-model} used throughout this paper.

\section{The dispersion-relation for a narrowband resonator\label{sec:dispersion-relation}}
In this section we will present a theoretical result from our framework. We will demonstrate that the dispersion-relation equations developed in previous works~\cite{Venturini2018,Lindberg2018,Cullinan2022} can be obtained from the Lebedev equation as special cases. With that we will prove the equivalence of the two approaches under certain conditions.

For the particular case of a narrowband resonator, we can retain a single harmonic $\pm p_0$ and $G_{pp^{\prime}}(\Omega)$ is a $2\times 2$ matrix. For this case, the Lebedev equation yields
\begin{align}
0 =&~\mathrm{det}\mqty[1 + i \kappa M_{-p_0-p_0}(\Omega) & i \kappa M_{-p_0p_0}(\Omega) \\ i \kappa M_{p_0-p_0}(\Omega)  & 1 + i \kappa M_{p_0p_0}(\Omega)] \nonumber \\
\approx&~1 + i\kappa \left(M_{p_0p_0}(\Omega) + M_{-p_0-p_0}(\Omega)\right),
\label{eq:determinant}
\end{align}
where $M_{pp^{\prime}} = \frac{Z_{p, \ell}(\Omega)}{\omega_{p, \ell}} G_{pp^{\prime}}(\Omega)$. The approximation refers to $M_{-p_0p_0}M_{p_0-p_0} \approx M_{p_0p_0}M_{-p_0-p_0}$, which follows from the property $H_{m, -p}(J) \approx H_{m, p}^{\ast}(J)$ that can be checked from Eq.~\eqref{eq:hmp}. The approximation is better for $p_0 h \gg \ell$.

We will assume symmetric elliptical orbits on the longitudinal phase-space, thus the canonical transformation can be approximately factored as
\begin{equation}
    \zeta(J, \varphi) \approx f(J) \cos(\varphi).
    \label{eq:canonical_case}
\end{equation}

This form is exact for a quadratic (harmonic) potential and a good approximation even for a quartic potential, as discussed in the Appendix B of Ref.~\cite{Lindberg2018}. With this form, we have that
\begin{align}
    H_{m, p}(J) &\approx \frac{1}{2\pi} \int_{0}^{2\pi} \dd{\varphi} e^{im\varphi + i\omega_{p, \ell} f(J)\cos(\varphi)/c} \nonumber \\
    &= i^{m} \mathcal{J}_m\left(\omega_{p, \ell} f(J)/c\right),
\end{align}
where $\mathcal{J}_m(x)$ is the Bessel function of the first kind.

We can use the approximation that the wakefield varies slowly over the length of the bunch. In Refs.~\cite{Lindberg2018, Cullinan2022}, this justifies a Taylor expansion of the longitudinal wake function keeping only low-order terms. In our framework, this limit corresponds to consider a short-bunch, $\omega_{p_0, \ell} f(J)/c \ll 1$, so we can use the approximation of the Bessel function for small arguments:
\begin{equation}
    \mathcal{J}_m\left(x\right) \approx i^m\frac{(x/2)^m}{m!}.
\end{equation}

From the relation between $\Psi_0(J)$ and $\mathcal{H}_0(J)$, we can show that $\pdv{\Psi_0}{J} = - \frac{\omega_s(J)}{\alpha c \sigma_\delta^2 }\Psi_0(J)$. Then, in the short-bunch limit, the determinant Eq.~\eqref{eq:determinant} results in:
\begin{align}
    1 &= \frac{2i \kappa}{\alpha c \sigma_\delta^2}  \sum_{p = \pm p_0}\frac{Z_{p,\ell}(\Omega)}{\omega_{p, \ell}} \int_{0}^{\infty} \dd{J}
    \Psi_0(J) \nonumber \\
    & \times \sum_{m=1}^{\infty} \frac{(\omega_{p, \ell}/c)^{2m}}{(m!)^2}\frac{m^2 \left[f(J)/2\right]^{2m}}{\left[\Omega/\omega_s(J)\right]^2 - m^2}.
\end{align}

Let us define the normalized effective impedance of order $n$:
\begin{equation}
    Z_{\eff, \ell}^{(n)}(\Omega) = \sum_{p=\pm p_0}\left(\sigma_z \omega_{p, \ell} / c\right)^n Z_{p, \ell}(\Omega).
    \label{eq:effective_impedance}
\end{equation}
Note that the factor $\sigma_z \omega_{p, \ell} / c$ is dimensionless, and it is worth mentioning that the bunch length $\sigma_z$ (in this work, always taken as the second central moment of $\lambda_0(z)$) was not fundamental, it was introduced only for the purpose of a convenient normalization. In the Appendix~\ref{appendix:impedance-wake} we show that the normalized impedance of order $n$ can be related to the $n$th derivative of the wake function.

With $\kappa$ given by Eq.~\eqref{eq:kappa}, the dispersion-relation for all azimuthals $m$ is:
\begin{align}
    1 &= i\frac{4\pi eI_0 \sigma_z}{E_0 C_0 \alpha \sigma_\delta^2 }  \int_{0}^{\infty} \dd{J}
    \Psi_0(J) \nonumber \\
    &\times \sum_{m=1}^{\infty}\frac{Z_{\eff, \ell}^{(2m-1)}(\Omega)}{(m!)^2}\frac{m^2 \left[f(J)/2\sigma_z\right]^{2m}}{\left[\Omega/\omega_s(J)\right]^2 - m^2}.
    \label{eq:dispersion_m}
\end{align}
We define the $\Lambda_\ell^{(m)} (\Omega)$ parameter as:
\begin{align}
    \Lambda_\ell^{(m)} (\Omega) &= i \frac{eI_0}{2E_0T_0 \sigma_\delta} \frac{Z_{\eff, \ell}^{(2m-1)}({\Omega})}{(m!)^2},
    \label{eq:lambda_imp_general}
\end{align}
and replacing this definition into Eq.~\eqref{eq:dispersion_m} yields:
\begin{equation}
        1 =  \frac{2 \sigma_z}{\alpha c \sigma_\delta} \sum_{m=1}^{\infty}\Lambda_\ell^{(m)}(\Omega) D_m(\Omega),
        \label{eq:dispersion_relation_zj}
\end{equation}
with the dispersion integral for the azimuthal mode $m$:
\begin{equation}
    D_m(\Omega) = \int_{0}^{\infty} \dd{J} 4 \pi\Psi_0(J)\frac{m^2\left[f(J)/2\sigma_z\right]^{2m}}{\left[\Omega/\omega_s(J)\right]^2 - m^2}.
    \label{eq:dispersion_integral}
\end{equation}

The result we have obtained in Eq.~\eqref{eq:dispersion_relation_zj} allows the evaluation of instabilities for arbitrary azimuthal modes independently or combined. The assumptions to achieve this were: (i) elliptical symmetric orbits in the phase-space and (ii) the wakefield length is much longer than the bunch length, i.e., the short-bunch limit.

The dispersion integral can be simplified to specific rf potentials, as done in Refs.~\cite{Krinsky1985, Venturini2018, Lindberg2018}.
\subsection{Dipole instabilities\label{subsec:dipole-instability}}
A particularly important instability type regards the bunch centroid motion, referred to as dipole instability. This can be studied by focusing on the $m=1$ mode contribution for the dispersion-relation.

As in Ref.~\cite{Lindberg2018}, the canonical transformation can be expanded in Fourier series $\zeta(J, \varphi) = \sum_{\nu}e^{i\nu\varphi}\hat{f}_\nu(J)$ and the Fourier coefficients $\hat{f}_\nu(J)$ appear in the dispersion-relation instead of the function $f(J)$. For the case $\zeta(J, \varphi) = f(J)\cos(\varphi)$, the only non-zero Fourier coefficients are $\nu=\pm 1$ and the result $\hat{f}_{\pm 1}(J) = f(J)/2$ can be used. Since the canonical transformation is real, $\hat{f}_{-1}(J) = \hat{f}_{1}(J)$. Under these conditions, the dispersion-relation Eq.~\eqref{eq:dispersion_relation_zj} for $m=1$ is:
\begin{equation}
    1 =  \frac{2 \sigma_z}{\alpha c \sigma_\delta}\Lambda_\ell^{(1)}(\Omega) \int_{0}^{\infty} \dd{J} 4 \pi\Psi_0(J)\frac{\left[\hat{f}_1(J)/\sigma_z\right]^{2}}{\left[\Omega/\omega_s(J)\right]^2 - 1}.
    \label{eq:dipole_dispersion_relation}
\end{equation}

To connect with Lindberg's approach presented in~\cite{Lindberg2018}, we will first solve the dispersion-relation for an equivalent harmonic problem, i.e., a quadratic potential $\Phi_0(z) \propto z^2$ producing the same bunch length $\sigma_z$ related to the arbitrary potential of interest. In this case, the following conditions apply:
\begin{align}
    \omega_s(J) &= \omega_{s0} = \alpha c \sigma_\delta / \sigma_z, \\
    \mathcal{H}_0(J) &= \omega_{s0} J / c, \\
    \Psi_0(J) &= \frac{e^{-J/\langle J \rangle}}{2\pi \langle J \rangle} \qq{with} \langle J \rangle = \sigma_z \sigma_\delta, \\
    \zeta(J, \varphi) &= \sigma_z \sqrt{\frac{2J}{\langle J \rangle}} \cos(\varphi).
\end{align}

We can initially obtain the coherent frequency $\Omega_{\mathrm{linear}, \ell}$ related to the centroid motion of the coupled-bunch mode $\ell$ with linear dynamics. Additionally, we will assume that we can approximate $\Omega \approx \langle\omega_s(J)\rangle = \omega_{s0}$ in the impedance's argument~\cite{Lindberg2018}. Applying these conditions we obtain:
\begin{align}
    D_1(\Omega_{\mathrm{linear, \ell}}) &= \frac{\omega_{s0}^2}{\Omega^2_{\mathrm{linear, \ell}} - \omega_{s0}^2}\label{eq:dispersion_d1}, \\
    \Lambda_\ell^{(1)}(\omega_{s0}) &= i \frac{eI_0}{2E_0T_0 \sigma_\delta} Z_{\eff, \ell}^{(1)}, \label{eq:lambda_imp} \\
    &=  \frac{-eI_0 \sigma_z}{2E_0 \sigma_\delta M} \nonumber \\
    &~~\times \sum_{k=0}^{+\infty} e^{i k (2\pi \ell + \omega_{s0}T_0)/M}\eval{\dv{W}{\xi}}_{\xi = k\frac{cT_0}{M}},\label{eq:lambda_wake}
\end{align}
where we used Eq.~\eqref{eq:wake_and_impedance} to relate $Z_{\eff, \ell}^{(1)}$ with the first derivative of the wake function.

Note that $\frac{eI_0 \sigma_z}{2E_0\sigma_\delta M} = \frac{e^2 \sigma_t N^\mathrm{part}}{2 \gamma m c T_0 \sigma_\delta}$ (for definition of the parameters in the right, see~\cite{Lindberg2018}), then we can show that Eq.~\eqref{eq:lambda_wake} is equivalent to the matrix elements in Eq.~(19) of~\cite{Lindberg2018}, after diagonalization to the basis of coupled-bunch modes. Combining Eqs.~\eqref{eq:dispersion_d1} and~\eqref{eq:lambda_imp} in the dispersion-relation, the result for point-charge bunches in a single-rf system is recovered:
\begin{align}
    \Omega_{\mathrm{linear, \ell}} &= \omega_{s0}\sqrt{1  + 2 \Lambda_\ell^{(1)}(\omega_{s0})/\omega_{s0}}.
\end{align}
For small detuning such that $\Lambda_\ell^{(1)}(\omega_{s0})/\omega_{s0} \ll 1$, we get Sacherer's formula $\Omega_{\mathrm{linear, \ell}} \approx \omega_{s0} + \Lambda_\ell^{(1)}(\omega_{s0})$. Note that $\Lambda_\ell^{(1)}(\omega_{s0})$ is actually the coherent frequency shift in a harmonic potential.

For comparison with Eq.~\eqref{eq:dipole_dispersion_relation}, see the dispersion-relation presented in Eq.~(24) of~\cite{Lindberg2018}, where the integral contains a summation over $m$. There, $\lambda_\ell$ is an eigenvalue of the bunch centroids coupling matrix, Eq.~(19) in~\cite{Lindberg2018}. These eigenvalues $\lambda_\ell$ are identical to the $\Lambda_\ell^{(1)}$ defined by Eq.~\eqref{eq:lambda_wake}, where the particularity to $m=1$ is made explicit in the notation. For $m=1$, our framework reproduces the results from Ref.~\cite{Lindberg2018}, considering that all the studies cases of rf potentials in that work considered only the $m=1$ contribution as well.

To obtain Eq.~(24) of Ref.~\cite{Lindberg2018} for arbitrary $m$ from Eq.~\eqref{eq:dispersion_relation_zj} of this paper, we would have to consider $\Omega \approx \omega_{s0}$ to evaluate $\Lambda^{(1)}_\ell(\omega_{s0})$, which is a reasonable approximation, in principle. Besides, we would also have to set $\Lambda^{(m)}_\ell = \Lambda^{(1)}_\ell$ for all azimuthals $m$ and assume that $\left[f(J)/2\right]^{2m} = \hat{f}_m^2(J)$, which are considerations that I could not find arguments to support. Therefore, for $m>1$, it was not possible to establish an obvious connection between our framework and the dispersion-relation in Ref.~\cite{Lindberg2018}.

\subsection{Quadrupole instabilities}
We will briefly address another type of instability to illustrate how the generality of the presented theory allows to straightforwardly obtain the dispersion-relation for any azimuthal mode. For instance, quadrupolar instabilities were investigated in Ref.~\cite{Cullinan2022}. Let us assume that $\Lambda_\ell^{(1)} = 0$, meaning that the dipole coherent shift is fully suppressed. Then, from Eq.~\eqref{eq:dispersion_relation_zj}, the next relevant contribution is from the quadrupole mode $m=2$.

Taking the normalized effective impedance of third order in terms of the wake function from Eq.~\eqref{eq:wake_and_impedance}, replacing $f(J)/2 = \hat{f}_1(J)$ in the dispersion integral, and rewriting $\Psi_0(J) = - \frac{\alpha c \sigma_\delta^2}{\omega_s(J)}\pdv{\Psi_0}{J}$, we get:
\begin{align}
    1 &= \frac{4 \pi eI_0}{E_0 M} \sum_{k=0}^{+\infty} e^{i k (2\pi \ell + \Omega T_0)/M }\eval{\dv[3]{W}{\xi}}_{\xi = k\frac{cT_0}{M}}  \nonumber \\
    &~~\times \int_{0}^{\infty} \dd{J}
    \frac{1}{\omega_s(J)}\pdv{\Psi_0}{J} \frac{\left[\hat{f}_1(J)\right]^{4}}{\left[\Omega/\omega_s(J)\right]^2 - 4}.
    \label{eq:quadrupole_dispersion_relation}
\end{align}
This dispersion-relation is equivalent to Eq.~(16) in Ref.~\cite{Cullinan2022}, assuming an even filling pattern and diagonalization to the coupled-bunch basis.

It is worth mentioning that derivations for dipole instabilities in~\cite{Lindberg2018}, later also adapted for quadrupole instabilities in~\cite{Cullinan2022}, required the restriction to these cases as initial assumptions for the theoretical development. This case-by-case approach may be limited if one wants to study an instability related to an azimuthal mode $m \notin \left\{1, 2 \right\}$ or if multiple $m$ modes are required to accurately compute the instability thresholds, for instance mode-coupling instabilities. Moreover, the mathematical complexity of the process increases with $m$, as evident from the comparison of Ref.~\cite{Cullinan2022} for $m=2$ with Ref.~\cite{Lindberg2018} for $m=1$.

Interestingly, the connection between an $m$th-order instability and the derivatives of odd orders $2m-1$ of the wake function naturally arises in our framework through Eqs.~\eqref{eq:lambda_imp_general} and~\eqref{eq:wake_and_impedance}. This aligns with the physical intuition that derivatives of even orders of the wake function cannot drive instabilities due to their symmetric effects.

\section{Applications\label{sec:applications}}
The developed theory will be applied to two instabilities of interest in recent publications, specially to some 4th-generation storage rings with \glspl{hc}, whose parameters are presented in Table~\ref{table:ring_params}. For the applications, we will benchmark the results from Lebedev equation, effective synchrotron frequency model and Gaussian \gls{lmci} against each other and against experimental data.
\begin{table*}
\caption{\label{table:ring_params}Relevant parameters for the longitudinal instability analysis of different 4th-generation storage rings.}
\begin{ruledtabular}
\begin{tabular}{lcccc}
    Parameter & Unit & ALS-U~\cite{Venturini2018} &  HALF~\cite{He2022a} & MAX IV~\cite{Cullinan2020IPAC, Cullinan2024}\\
    \hline
    Energy~$E_0$ & $\SI{}{\giga\electronvolt}$                            & \num{2.0}     & \num{2.2} & \num{3.0} \\
    Beam current (uniform fill)~$I_0$ & $\SI{}{\milli\ampere}$                        & \num{500}     & \num{350} & \numrange{200}{400} \\
    Circumference~$C_0$ & $\SI{}{\meter}$                                 & \num{196.5}   & \num{480.0} & \num{528.0} \\
    Harmonic number~$h$ &                                               & \num{328}     & \num{800}   & \num{176} \\
    Momentum compaction factor~$\alpha$ & $\SI{}{}$                          & \num{2.11e-4} & \num{8.1e-5} & \num{3.06e-4}\\
    Energy loss per turn~$U_0$ & $\SI{}{\kilo\electronvolt}$              & \num{217}  & \num{198.8} & \num{363.8} \\
    Relative energy spread~$\sigma_\delta$ & $\SI{}{}$                              & \num{9.43e-4} & \num{6.43e-4} & \num{7.69e-4} \\
    Natural std bunch length~$\sigma_z~(\sigma_\tau)$ & $\SI{}{\milli\meter}~(\SI{}{\pico\second})$  & \num{3.5}~(11.8)    & \num{2.0}~(\num{6.8}) & \num{10.9}~(\num{36.4}) to \num{12.1}~(\num{40.4})\\
    Radiation damping time~$\tau_\delta$ & $\SI{}{\milli\second}$ &\num{14.0} & \num{22.7} & \num{25.2}\\
    rf frequency~$f_\rf$ &  $\SI{}{\mega\hertz}$                            & \num{500.417} & \num{499.654} & \num{99.931} \\
    Revolution frequency~$f_0$ &  $\SI{}{\kilo\hertz}$                      & \num{1525.66} & \num{624.57} & \num{567.69} \\
    \glspl*{MC} total voltage~$\hat{V}_\rf$ & $\SI{}{\mega\volt}$                 & \num{0.6} & \num{0.85} &\numrange{1.0}{1.2} \\
    \gls*{hc} technology  &                                        & NC & SC & NC \\
    \gls*{hc} rf harmonic &                                        & \num{3} & \num{3} & \num{3} \\
    Number of \gls*{hc}s  &                                        & \num{2} & \num{1} & \numrange{2}{3} \\
    \gls*{hc} shunt impedance~$R_s = V^2/2P$ & $\SI{}{\mega\ohm}$     & \num{1.7} & \num{45} & \num{2.75} \\
    \gls*{hc} quality factor~$Q$ & $\SI{}{}$                           & \num{2.1e4} & \num{5e5} & \num{2.08e4} \\
    \gls*{hc} geometric factor~$(R/Q)$ & $\SI{}{\ohm}$                  & \num{81} & \num{90} & \num{132} \\
    \gls*{hc} flat-potential voltage~$\hat{V}_\hc$  & $\SI{}{\mega\volt}$        &  \num{186.6} & \num{283.3} & \numrange{307.5}{378.6} \\
    \gls*{hc} flat-potential detuning~$\Delta f_\hc$ & $\SI{}{\kilo\hertz}$       & \num{584}   & \num{157.8} & \numrange{38.8}{145.2}\\
\end{tabular}
\end{ruledtabular}
\end{table*}

\subsection{Robinson dipole-quadrupole mode coupling}
Robinson instabilities can be studied by focusing on the coupled-bunch mode $\ell=0$. In particular, there is a Robinson instability caused by the coupling of the dipole and quadrupole motion, driven by the \gls{hc} impedance, that has been studied in simulations~\cite{Bosch2001,Gamelin2024} and observed experimentally at MAX IV~\cite{Cullinan2020IPAC, Cullinan2024}.

Figure~\ref{fig:dipole-quadrupole-coupling} shows the coherent frequencies calculated with different methods and the measured values at MAX IV ring~\cite{Cullinan2020IPAC}. The low total current of $\SI{50}{\milli\ampere}$ allowed to measure the coherent oscillation frequencies with a stable beam. The incoherent effective synchrotron frequency calculated by Eq.~\eqref{eq:freq_by_bunlen} is also shown to indicate its reduction while the \gls{hc} voltage increases. In contrast, the coherent dipolar frequency for the $\ell=0$ mode remains essentially constant and equal to the value of the single-rf system (for a physical explanation, Ref.~\cite{Ng2006}, pages 68 and 206). The coherent quadrupolar frequency follows the reduction of the second harmonic of the incoherent frequency. Hence, at some \gls{hc} voltage the dipole and quadrupole modes will match. For low currents such as $I_0 = \SI{50}{\milli\ampere}$, the modes actually only cross each other and do not couple to drive an instability. For higher currents these modes typically couple, driving a fast instability that can prevent reaching higher \gls{hc} voltages, thus better bunch lengthening performance.
\begin{figure}
    \includegraphics[width=0.48\textwidth]{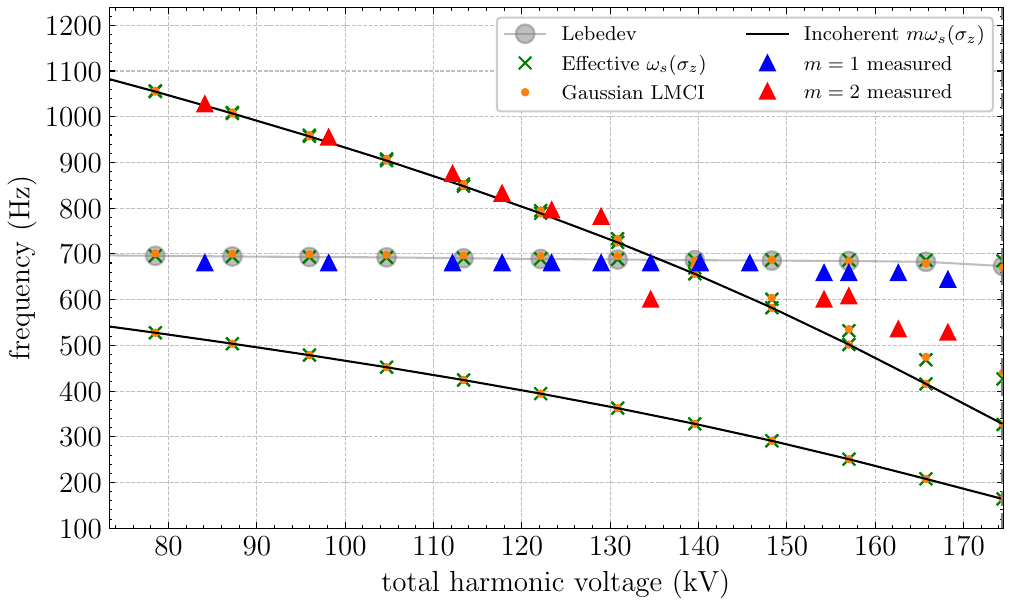}
    \caption{\label{fig:dipole-quadrupole-coupling} Robinson ($\ell=0$) dipole-quadrupole mode coupling for MAX IV parameters: $I_0 = \SI{50}{\milli\ampere}$, $\hat{V}_\rf=\SI{650}{\kilo\volt}$. The flat-potential harmonic voltage is \SI{174.35}{\kilo\volt}. Measured data from MAX IV taken from Ref.~\cite{Cullinan2020IPAC}. All models predict no instability at this condition, in agreement with the experiment. $m_\mymax=2$ was used in all methods and $k_\mymax=1$ was used in Gaussian LMCI.}
\end{figure}

It is interesting to note that all models produced very similar results. In this condition, solutions with $\Im(\Omega) < 1/\tau_\delta$ where found for the Lebedev equation, and other solutions following the quadrupole frequencies could also be found if the initial guess to solve the determinant root was chosen to be close to the second harmonic of the incoherent frequency. In the case of instability, however, an initial guess around the dipole frequency would be sufficient to correctly predict an unstable solution.

The good agreement between all methods and the measured data indicates that the contributions from a non-Gaussian bunch and nonlinearities in the double-rf system are not important for determining the coherent frequencies. It was observed that this still holds for predicting an unstable condition. Thus, only the effects in the bunch length and synchrotron frequency as in an equivalent single-rf system proved to be sufficient to study the Robinson mode coupling instability. This aligns with the observations from previous investigations~\cite{Bosch2001,Cullinan2020IPAC}. Benchmarking of the Gaussian \gls{lmci} model with tracking simulations are reported elsewhere~\cite{Gamelin2024}.
\subsection{PTBL/mode-1 instability}
The \gls{ptbl} instability, also called by some authors as mode-1 instability, has been recently investigated for 4th-generation storage rings with \glspl{hc}~\cite{Venturini2018, He2022a, Cullinan2024}. In this paper, \gls{ptbl} or mode-1 instability refer to the same effect. It was shown that during this instability, the bunches centroids and profiles oscillate in a quasi-stationary motion. Some studies indicate that the effect has different features from a standard coupled-bunch instability~\cite{He2022a,Gamelin2022}, yet some success was obtained in computing the thresholds by restricting the analysis to the coupled-bunch mode $\ell=1$ as it is the dominant mode. This is the justification for the \enquote{mode-1 instability} name.

In Ref.~\cite{He2022a}, the characteristics of \gls{ptbl} were investigated in detail mainly through tracking simulations, although discussions on the instability mechanism were not addressed. In Ref.~\cite{Cullinan2024} the nature of this instability was further explored and the authors elaborated on some conditions that should be met for the mode-1 instability be likely to appear. The experimental data obtained at MAX IV \SI{3}{\giga\electronvolt} storage ring (see Fig. 11 in Ref.~\cite{Cullinan2024}) showed a significant disagreement when compared to results obtained by two theoretical models: Krinsky dispersion-relation for a quartic potential, in the format presented in~\cite{Lindberg2018}, and T. He formula~\cite{He2022b}. We will present the results obtained from the models developed in this paper, from which we could obtain new insights to understand what features are important to predict~\gls{ptbl} and why previous theories were unsuccessful in some cases.

Figure~\ref{fig:ptbl-rings} shows the coherent frequencies of the mode-1 calculated by different methods for different rings, with the unstable region indicated by the background red color. We note that for ALS-U parameters using the \enquote{old} ALS \glspl{hc}, the mode-1 is unstable for all \gls{hc} conditions, in accordance with the results presented in Ref.~\cite{Venturini2018}. This motivated a new design of \glspl{hc} for ALS-U. For HALF parameters with \SI{350}{\milli\ampere}, the mode-1 instability is triggered when the \gls{hc} voltage is \SI{6}{\percent} below the flat-potential voltage. This aligns with the results from Ref.~\cite{He2022a}, reporting a \gls{ptbl} threshold of \SI{259}{\milli\ampere} when the \gls{hc} is at flat-potential. Finally, for MAX IV parameters with \SI{300}{\milli\ampere} and 3 \glspl{hc}, the mode-1 instability is driven \SI{1}{\percent} below the flat-potential voltage.
\begin{figure}
    \centering
    \subfloat[ALS-U.\label{ptbl-als-u}]{
    \includegraphics[width=0.48\textwidth]{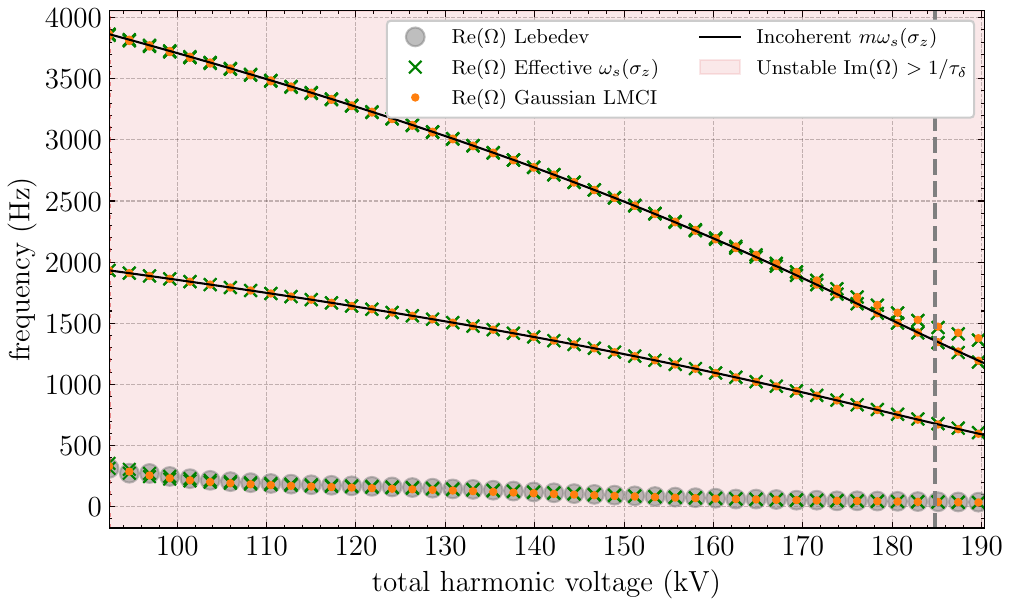}
    }
    \hfill
    \subfloat[HALF.\label{ptbl-half}]{
    \includegraphics[width=0.48\textwidth]{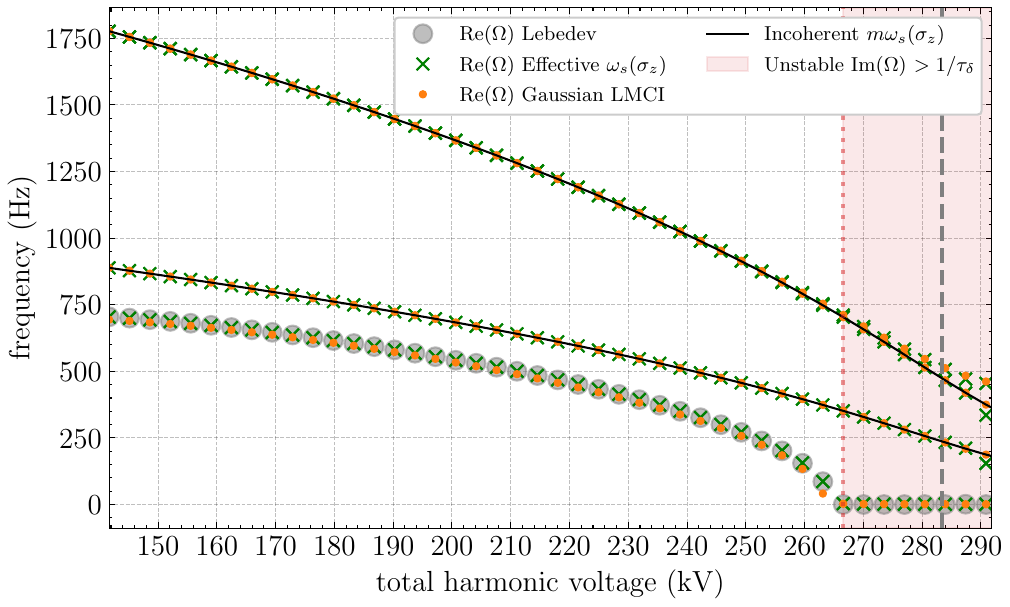}
    }
    \hfill
    \subfloat[MAX IV.\label{ptbl-max-iv}]{
    \includegraphics[width=0.48\textwidth]{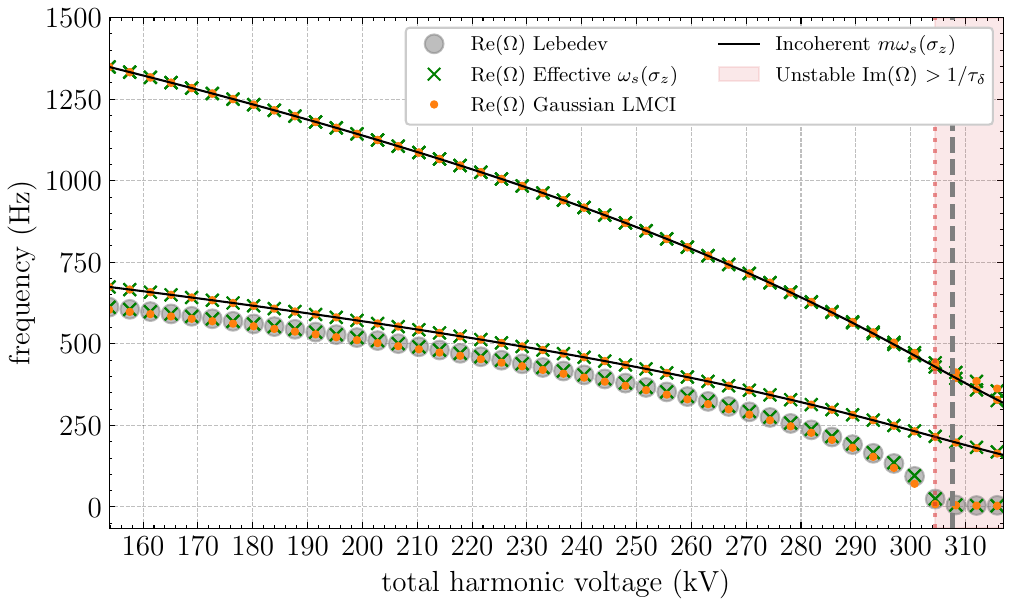}
    }
    \caption{Coherent frequencies of coupled-bunch mode $\ell=1$ as a function of the \gls{hc} voltage for different storage ring parameters. The vertical gray dashed line indicates the flat-potential voltage. The unstable area is determined by the condition $\Im(\Omega) > 1/\tau_\delta$ with $\Omega$ being the solution of Lebedev equation. (a) Unstable for all \gls{hc} voltages. flat-potential \SI{184.75}{\kilo\volt}. (b) Threshold \SI{266.58}{\kilo\volt}. Flat-potential \SI{283.35}{\kilo\volt}. (c) Threshold \SI{304.48}{\kilo\volt}. Flat-potential \SI{307.62}{\kilo\volt}. $I_0=\SI{300}{\milli\ampere}$, $\hat{V}_\rf=\SI{1.0}{\mega\volt}$, 3~\glspl{hc}. $m_\mymax=2$ was used in all methods and $k_\mymax=1$ was used in Gaussian LMCI.\label{fig:ptbl-rings}}
\end{figure}

The results from Fig.~\ref{fig:ptbl-rings} also reveal that, for the mode-1 instability, calculations with a more complete theory (Lebedev equation) produce essentially the same values as calculations with theories that neglects the Landau damping effect (effective synchrotron frequency and Gaussian \gls{lmci}). This is a strong evidence that Landau damping does not play a role on the onset of the \gls{ptbl} instability, contrary to the conclusions from Ref.~\cite{Cullinan2024}. Another relevant observation from Fig.~\ref{fig:ptbl-rings} is that the low coherent frequency feature of \gls{ptbl} instability was reproduced. In fact, from HALF and MAX IV plots, it can be understood that the coherent frequency is shifted to lower values more than the incoherent frequency, until a point it approaches zero and the instability is triggered. With this picture, the mechanism of the instability associated with the imaginary/reactive part of the impedance can be better understood, because this term is responsible for frequency shifts. In Fig.~\ref{fig:measured-mode-1} we benchmarked our predictions with the coherent frequencies of mode-1 measured at MAX IV~\cite{Cullinan2024}, displaying very good agreement as well. The measurements were carried out at \SI{90}{\milli\ampere}, where the mode-1 instability is not triggered. We see that for such low current the coherent frequency shift is negligible.
\begin{figure}
    \includegraphics[width=0.48\textwidth]{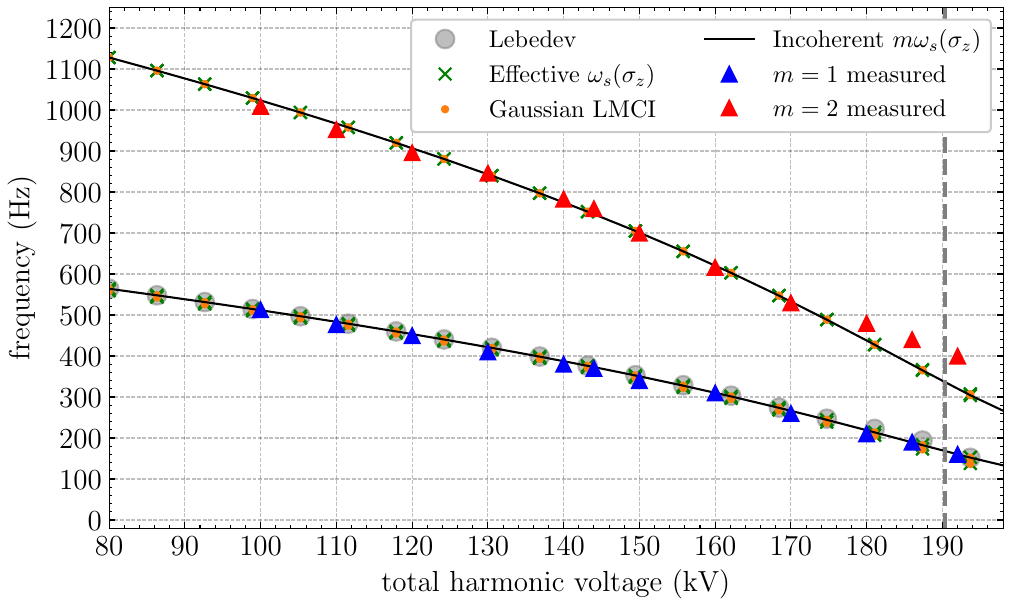}
    \caption{Coherent frequencies of coupled-bunch mode $\ell=1$ for MAX IV parameters with 2 \glspl{hc}, $I_0 = \SI{90}{\milli\ampere}$, $\hat{V}_\rf=\SI{689}{\kilo\volt}$. Comparison between different calculation methods and experimental data (Fig.~15 from Ref.~\cite{Cullinan2024}).\label{fig:measured-mode-1}}
\end{figure}
\begin{figure}
    \includegraphics[width=0.48\textwidth]{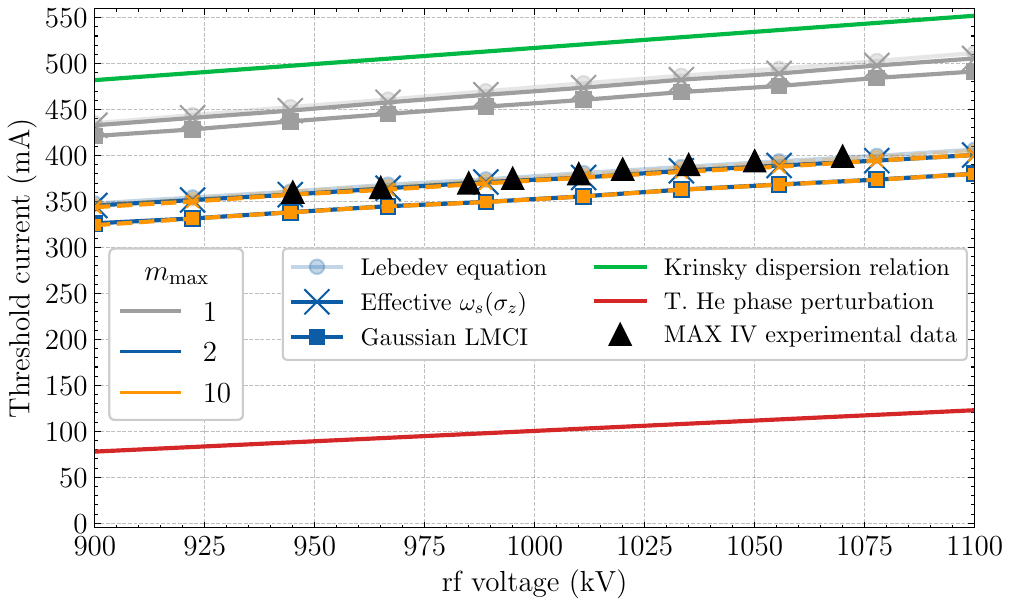}
    \caption{Threshold currents of the mode-1 instability for different main rf voltages. MAX IV parameters with 2 \glspl{hc}, tuned to provide the flat-potential voltage. Comparison between different calculation methods and experimental data~\cite{Cullinan2024}. Krinsky and T. He curves were obtained from Fig.~11 in~\cite{Cullinan2024}. The truncation of azimuthal modes $m_\mymax$ was varied to illustrate the relevance of multiple modes. $k_{\mymax} =1$ was used in Gaussian LMCI.\label{fig:ptbl-vs-rfvoltage}}
\end{figure}

An interesting feature, explored in simulations~\cite{He2022a} and measured at MAX IV~\cite{Cullinan2024}, is the  dependence of the threshold current with the main rf voltage, which shows a linear trend with positive slope. In the experiments, the \gls{hc} cavity was always tuned to produce the flat-potential voltage for each current and rf voltage. We benchmarked our calculations with the experimental data from MAX IV and results from other methods as reported in Ref.~\cite{Cullinan2024}. The comparison is shown in Fig.~\ref{fig:ptbl-vs-rfvoltage}. The results obtained from the Lebedev equation and the effective synchrotron frequency model are in excellent agreement with the experimental data when $m_\mymax\geq2$ azimuthal modes are accounted. The Gaussian \gls{lmci} model systematically predicts a lower current threshold, but still much more accurate than the Krinsky and T. He models. The agreement between the result from Lebedev equation and the effective synchrotron frequency reveals that the effects of Landau damping are not necessary to accurately predict the PTBL/mode-1 instability threshold. 

It was proven that the inclusion of the $m=2$ mode is essential to reproduce the measured thresholds. The calculations with only $m=1$ predicts a higher threshold such as in the Krinsky model (which only uses $m=1$). A more detailed discussion about the Landau damping and multiple azimuthal modes will be addressed in Sec.~\ref{sec:discussion}. Besides, the effects of a non-Gaussian beam proved to be relevant because it is the main difference between the effective frequency method and the Gaussian \gls{lmci} scheme, with the latter underestimating the threshold. The inclusion of multiple azimuthal modes is one of the main differences from our theoretical models to the models of Krinsky and T. He.
\section{Discussion on PTBL/mode-1 mechanism\label{sec:discussion}}
The results in Sec.~\ref{sec:applications} are helpful to understand the underlying mechanism of the \gls{ptbl} instability. For increasing voltage in a passive \gls{hc}, $\langle\omega_s\rangle$ reduces from its single-rf value. For the $\ell=0$ coupled-bunch mode, the coherent dipole frequency remains constant. This does not hold in general for other coupled-bunch modes and their coherent frequencies are expected to vary with the passive \gls{hc} fields. Figure ~\ref{fig:ptbl-rings} shows that, particularly for the $\ell=1$ mode, the coherent frequencies represented by \enquote{$\Re(\Omega)$ Lebedev} have a negative shift in relation to $\langle\omega_s\rangle$. This is due to the reactive (imaginary) part of the \gls{hc} impedance. If the negative shift leads to $\Re(\Omega) \approx 0$, the instability is triggered due to the lack of focusing. 

The zero-frequency condition can be used to derive an approximate formula for the \gls{ptbl} instability threshold (see Appendix~\ref{appendix:formula}). The formula provides a critical $(R/Q) I_0$ value and a mode-1 instability is expected when this value is exceeded. This dependence aligns with previous studies, which have shown that \glspl{hc} with low $(R/Q)$ are preferable for avoiding the \gls{ptbl} instability~\cite{Venturini2018,Venturini2018b,He2022b,He2022b,Gamelin2024}. Within the proposed instability mechanism, this behavior can be attributed to the lower $(R/Q)$ values reducing the reactive effective impedance for the $\ell=1$ mode, which in turn reduces the coherent dipole frequency shift.

The linear dependence of the \gls{ptbl} threshold on the main rf voltage was predicted by tracking simulations~\cite{He2022a}, verified experimentally~\cite{Cullinan2024}, and reproduced with our calculations in Fig.~\ref{fig:ptbl-vs-rfvoltage}. In Fig.~\ref{fig:ptbl-vs-rfvoltage-half}, we show the behavior of the incoherent and coherent frequencies for two different main rf voltages, using the HALF parameters. For simplicity, only the results from the Lebedev equation are presented. The result in red corresponds to the condition shown in Fig.~\ref{fig:ptbl-rings}b, with $\hat{V}_\rf = \SI{0.85}{\mega\volt}$. This is compared with a result obtained at twice the rf voltage, $\SI{1.70}{\mega\volt}$, where the single-rf synchrotron frequency is expected to increase by approximately $\sqrt{2}$. At the higher rf voltage, the coherent negative shift is reduced. This reduction occurs due to the lower \gls{hc} detuning needed to generate a higher harmonic voltage, which decreases the reactive effective impedance for $\ell=1$. Combined with the higher incoherent frequency, this implies in an increase in the \gls{ptbl} threshold current. According to the approximate formula, Eq.~\eqref{eq:threshold-formula}, the threshold depends on $\sqrt{\hat{V}_\rf}$, while a linear behavior was observed in simulations~\cite{He2022a} and measurements~\cite{Cullinan2024}. We can argue that variations considerably larger than those made in these investigations would be required to reveal a $\sqrt{\hat{V}_\rf}$ dependence.
\begin{figure}
    \includegraphics[width=0.48\textwidth]{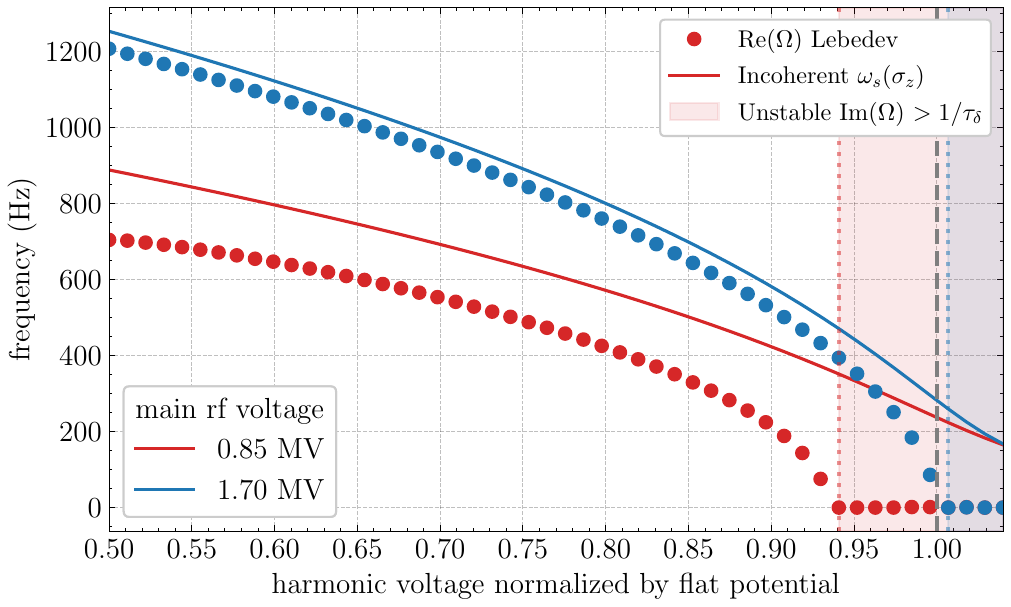}
    \caption{\label{fig:ptbl-vs-rfvoltage-half} Coherent frequencies of $\ell=1$ coupled-bunch mode for HALF parameters with different main rf voltages. Harmonic voltage/detuning at flat-potential for each rf voltage are (red): $\SI{283.35}{\kilo\volt}\slash \SI{157.79}{\kilo\hertz}$; (blue): $\SI{566.68}{\kilo\volt}\slash \SI{80.17}{\kilo\hertz}$. $m_\mymax=2$ was used.}
\end{figure}

For all cases studied here, the positive growth rates for \gls{ptbl} are on the order of hundreds of \SI{}{\hertz} or higher. Such large growth rates are typical of mode coupling instabilities, for which radiation damping is known to be ineffective~\cite{Chao1993,Ng2006,Suzuki1983}. This observation aligns with the findings in Ref.~\cite{He2022a}, where tracking simulations indicated that the \gls{ptbl} threshold is insensitive to changes in the radiation damping time. Additionally, recent studies have showed that a resistive feedback is ineffective in mitigating the mode-1 instability~\cite{Lindberg2023}. Investigating the effect of reactive feedback to counteract the negative coherent frequency shift of mode-1 could offer a potential solution to the \gls{ptbl} issue. Reactive feedback systems have previously been explored to increase the thresholds of transverse mode coupling instabilities, achieving positive results~\cite{Ruth1983a,Ruth1983b,Myers1985,Myers1987,Balewski1990}.

In Refs.~\cite{Cullinan2024,Gamelin2024,Alves2024HarmonLIP,Alves2024IPAC} it was remarked that, since the \gls{ptbl} instability is known to have a low coherent frequency, it may be resistant to Landau damping. In addition to the presence of an incoherent frequency spread, Landau damping requires an overlap between coherent and incoherent frequencies to manifest. The argument is that, although double-rf systems significantly increase the frequency spread, the bandwidth may not extend to the very low frequencies involved in the \gls{ptbl} instability, limiting the effectiveness of Landau damping. Our results provide quantitative support for this argument.

The Krinsky dispersion-relation used in Refs.~\cite{Lindberg2018, Cullinan2024} assumes an ideal quartic rf potential, $\Phi_0(z) \propto z^4$, resulting in an amplitude-dependent incoherent frequency, $\omega_s(J) \propto J^{1/3}$, and isolates the $m=1$ contribution \cite{Krinsky1985, Venturini2018, Lindberg2018}. However, achieving in practice an exact quartic potential with a double-rf system is unlikely. Even small mismatches in the rf voltage cancellation can significantly alter the potential (see Fig.~3 in Ref.~\cite{Venturini2018}, for example), leading to incoherent frequencies that may not reach zero to interact with the coherent frequency. Consequently, the Krinsky model is expected to overestimate Landau damping effects. Combined with the neglect of higher-order $m$ modes, this may explain the discrepancy with the measured mode-1 thresholds shown in Fig.~\ref{fig:ptbl-vs-rfvoltage}. It is worth noting that the dispersion-relation applied in Refs.~\cite{Lindberg2018, Cullinan2024} is a specific case of the broader framework introduced in Krinsky's original work~\cite{Krinsky1985}, which is also general enough to include multiple azimuthal modes and arbitrary nonlinear rf potentials.

Finally, we present the $m=2$ mode contribution to the \gls{ptbl} instability prediction. Figure~\ref{fig:ptbl-vs-m-modes-maxiv} illustrates how the number of azimuthal modes affects the coherent frequencies. The calculations use MAX IV parameters with a beam current of \SI{400}{\milli\ampere}, a main rf voltage of \SI{1.0}{\mega\volt}, and two \glspl{hc}, a condition known to be unstable (see Fig.~\ref{fig:ptbl-vs-rfvoltage}). Although no direct mode coupling mechanism triggers the instability, interactions between azimuthal modes play a crucial role. For this beam current, considering only $m=1$ results in an insufficient coherent shift to push the mode towards zero and drive the instability. Including $m=2$ introduces an additional negative shift, as if the quadrupole mode \enquote{repels} the dipole mode, which is enough to drive the instability.
\begin{figure}
    \includegraphics[width=0.45\textwidth]{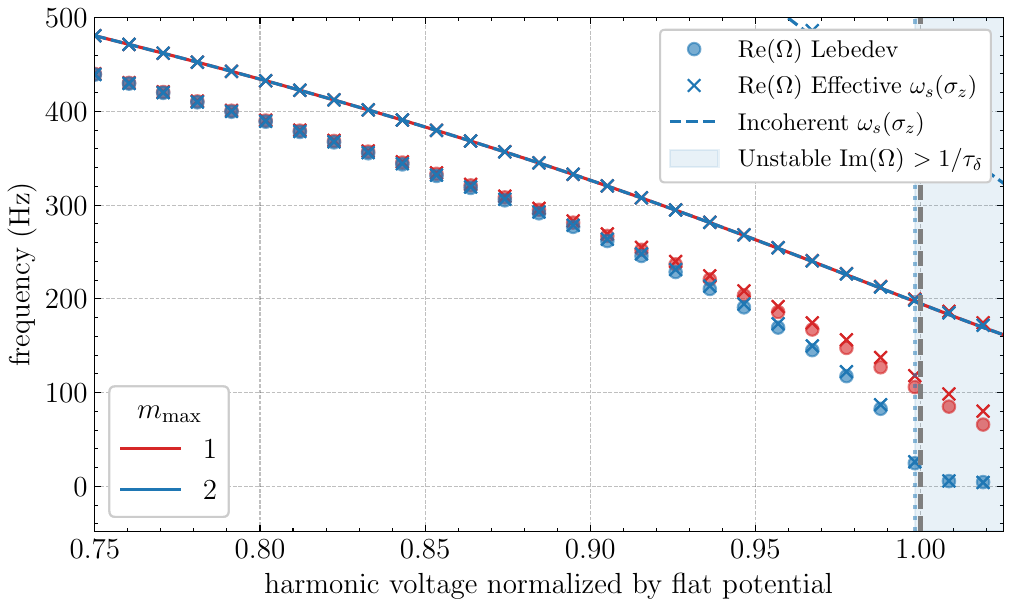}
    \caption{\label{fig:ptbl-vs-m-modes-maxiv} Coherent frequencies of $\ell=1$ coupled-bunch mode for MAX IV parameters with \SI{400}{\milli\ampere}, main rf voltage $\SI{1.0}{\mega\volt}$ and 2 HCs, considering two truncation values of azimuthal modes $m_\mymax$.}
\end{figure}

\section{Summary and conclusions\label{sec:conclusion}}
We developed a theoretical framework for calculating coupled-bunch instabilities in double-rf systems with \glspl{hc}, considering nonlinear effects from potential-well distortion and cases where multiple azimuthal modes are relevant. This framework is based on a frequency-domain perturbation theory to solve the linearized Vlasov equation, resulting in the Lebedev equation~\cite{Lebedev1968,Karpov2021}, which provides the coherent frequencies of the beam. We identified an equivalence between the Lebedev equation and the theory developed by Venturini~\cite{Venturini2018}, noting that Venturini's method has an avoidable computational complexity that increases significantly with the number of azimuthal modes considered.

Additionally, we presented two approximate models: an effective synchrotron frequency method, which neglects Landau damping but accounts for other effects of an arbitrary rf potential, and a Gaussian \gls{lmci} scheme adapted for double-rf systems. Dispersion-relation equations based on Krinsky's work~\cite{Krinsky1985}, as presented in recent publications by Lindberg~\cite{Lindberg2018} and Cullinan~\cite{Cullinan2022}, were derived as particular cases of the Lebedev equation. Altogether, this demonstrates an unification of recent theories addressing longitudinal instabilities in double-rf systems.

The framework was applied to study two types of instabilities in the presence of \glspl{hc}: Robinson dipole-quadrupole mode coupling and \gls{ptbl}/mode-1. For these studies, we used parameters from the storage rings ALS-U, HALF, and MAX IV. The theory provided excellent agreement with experimental data from MAX IV, a novel result for the mode-1 instability. We drew three significant conclusions about the \gls{ptbl} instability mechanism: (i) it is a \enquote{zero-frequency} instability for the coupled-bunch mode $\ell=1$, (ii) Landau damping is irrelevant for determining instability thresholds and (iii) the interaction of multiple azimuthal modes is the fundamental effect for accurate threshold predictions. The new insights on the \gls{ptbl} instability mechanism deepens our understanding of its behavior and dependence on parameters such as the main rf voltage, $(R/Q)$ of the \gls{hc}, reactive impedance, and longitudinal radiation damping time.

The Python implementation of the models is available in the open-source package \texttt{pycolleff}~\cite{pycolleff}, providing a useful semi-analytical tool for studying instabilities in electron storage rings with \gls{hc} systems.

Interesting directions for future research would be extending the framework to evaluate instabilities with uneven filling patterns and broadband resonators with a reasonable computational complexity, as well as investigating the use of reactive feedback to control the negative coherent frequency shift of the coupled-bunch mode $\ell=1$ in double-rf systems, aiming to increase the \gls{ptbl} instability thresholds.

\begin{acknowledgments}
    The author thanks F. H. de Sá (LNLS) for many useful discussions and for proofreading the manuscript.
\end{acknowledgments}
\appendix
\section{Effective impedance and wake function derivative\label{appendix:impedance-wake}}

The longitudinal wake function is related to the longitudinal impedance by:
\begin{equation}
    {W}({\xi}) = \frac{1}{2\pi}\int_{-\infty}^{\infty} \dd{\omega} Z(\omega) e^{-i \omega \xi/c},
\end{equation}
and it is straightforward to compute the $n$th derivative of the wake function:
\begin{equation}
    \dv[n]{W}{\xi} = \frac{(-i)^n}{2\pi c^{n}}\int_{-\infty}^{\infty} \dd{\omega} \omega^n Z(\omega) e^{-i \omega  \xi/c}.
    \label{eq:derivative-wake}
\end{equation}

Considering $M$ bunches evenly distributed, we will evaluate the wake function at the harmonics $kcT_0/M$. Then multiply it by $e^{i k(2\pi \ell + \Omega T_0)/M}$, where $\ell$ is the coupled-bunch mode. To apply the Poisson formula:
\begin{equation}
\sum_{k=-\infty}^{+\infty}e^{ik\omega T_0/M}= M\omega_0 \sum_{p=-\infty}^{+\infty}\delta(\omega - pM\omega_0),
\end{equation}
with $\omega_0 = 2\pi / T_0$, we will assume the integrand in Eq.~\eqref{eq:derivative-wake} does not diverge, allowing to interchange the summation with the integral. This is a valid assumption for narrowband resonators at low frequencies, when the impedance is well-represented by a single or few harmonics $p$, and it can be neglected elsewhere~\cite{Venturini2018, Karpov2024}.

After these steps, we obtain:
\begin{align}
    & \sum_{k=-\infty}^{+\infty} e^{i k (2\pi\ell + \Omega T_0)/M}\eval{\dv[n]{W}{\xi}}_{\xi = k \frac{cT_0}{M}} \nonumber \\
    &~= \frac{M(-i)^n}{c^{n}T_0 }\sum_{p=-\infty}^{+\infty} (\omega_{p, \ell} + \Omega)^n Z_{p, \ell}(\Omega),
    \label{eq:relation}
\end{align}
with $\omega_{p, \ell} = (pM + \ell) \omega_0$. Typically $\Re(\Omega) \ll \omega_0$, so we can approximate $(\omega_{p, \ell} + \Omega)^n\approx\omega_{p, \ell}^n$. Note that, for generality, the $\Omega$ dependence should be kept in the impedance's argument.

Considering the impedance can be neglected except at the harmonic $\pm p_0$, the definition of normalized effective impedance of order $n$ from Eq.~\eqref{eq:effective_impedance} can be applied into Eq.~\eqref{eq:relation} to get:
\begin{equation}
    Z_{\eff,\ell}^{(n)}(\Omega) = (i\sigma_z)^n\frac{T_0}{M} \sum_{k=0}^{+\infty} e^{i k (2\pi\ell + \Omega T_0)/M} \eval{\dv[n]{W}{\xi}}_{\xi = k\frac{cT_0}{M}},
    \label{eq:wake_and_impedance}
\end{equation}
where the causality $W(z<0)=0$ was used to restrict the sum for $k>0$.

\section{Approximate formulas for the PTBL/mode-1 threshold\label{appendix:formula}}
We will assume an even filling pattern with all buckets filled, $M=h$. Consider the longitudinal impedance resonator model:
\begin{equation}
    Z(\omega) = \frac{R_s}{1 + iQ \left(\frac{\omega_\mathrm{r}}{\omega} - \frac{\omega}{\omega_\mathrm{r}}\right)},
\end{equation}
where $R_s$ is the shunt impedance, $Q$ the quality factor and $\omega_\mathrm{r}$ the resonant frequency. For narrowband resonators with high-$Q$, we can approximate the reactive impedance by:
\begin{equation}
    \Im\left[Z(\omega)\right] \approx - \left(\frac{R}{Q}\right)\left(\frac{\omega_\mathrm{r}}{\omega} - \frac{\omega}{\omega_\mathrm{r}}\right)^{-1}.\label{eq:reactive-imp}
\end{equation}

We will assume that the resonator is detuned above the $n$th rf harmonic, $\omega_\mathrm{r} = n h \omega_0 + \Delta \omega$, representing the case of a $n$th-\gls{hc}. From Eq.~\eqref{eq:lambda_imp_general} for the dipole mode, $\Re\left[\Lambda_\ell^{(1)}(0)\right] \propto -\Im\left[Z_{\eff, \ell}^{(1)}(0)\right]$. We will be interested in the limit $\Omega_\ell \approx 0$, so the $\Omega_\ell$ sampling on the impedance was neglected. Evaluating the reactive effective impedance with $p_0 = n$ on Eq.~\eqref{eq:reactive-imp}, yields:
\begin{equation}
    \Im\left[Z_{\eff, \ell}^{(1)}(0)\right] \approx \sigma_\tau \left(\frac{R}{Q}\right) (nh\omega_0)^2 \frac{\Delta \omega}{(\ell\omega_0)^2 - \Delta \omega^2}\label{eq:reactive-effective-impedance}.
\end{equation}
We considered $nh \gg \ell$ and $nh\omega_0 \gg \Delta \omega$. \glspl{hc} typically operate at small detunings, $0 < \Delta \omega < \omega_0$ (see Table~\ref{table:ring_params}), so, for $\ell \neq 0$, $(\ell\omega_0)^2 - \Delta \omega^2 \approx (\ell\omega_0)^2$ is often a valid approximation for the flat-potential detuning. In this approximation, the bunch length, $\sigma_\tau$, is treated as an independent parameter. Even so, the value of $\sigma_\tau$ used in the formula should be consistent with the equilibrium bunch distribution for each condition.

We can show that the \gls{hc} peak voltage can be approximated by:
\begin{equation}
    \hat{V}_\hc \approx I_0 F_n \left(\frac{R}{Q}\right) \frac{nh\omega_0}{\Delta \omega},\label{eq:peak-hc-voltage}
\end{equation}
where $F_n$ is the real form factor at the $n$th rf harmonic~\cite{Tavares2014,Alves2023}. We are under the approximation of symmetric bunches, so the imaginary part of $F_n$ is zero and also used $Q\gg 1$ to obtain Eq.~\eqref{eq:peak-hc-voltage}. The \gls{hc} amplitude to produce the flat-potential is~\cite{Tavares2014}:
\begin{equation}
    \hat{V}_{\hc, \mathrm{flat}} = \frac{\hat{V}_\rf}{n} \sqrt{1 - \frac{n^2}{n^2-1}\left(\frac{U_0}{e\hat{V}_\rf}\right)^2} \approx \frac{\hat{V}_\rf}{n}.\label{eq:hc-flat-voltage}
\end{equation}
Combining these results and applying to Eq.~\eqref{eq:lambda_imp_general}, yields:
\begin{equation}
\Re\left[\Lambda_{\ell}^{(1)} (0)\right]  \approx - \frac{\pi e\sigma_\tau F_n n^4 h^3}{E_0T_0^2 \sigma_\delta\hat{V}_\rf \ell^2 } \left[I_0\left(\frac{R}{Q}\right) \right]^2.
\end{equation}
Note that $\Re\left[\Lambda_{\ell}^{(1)}\right] \propto -1/\ell^2$, meaning the most significant negative shift occurs for the $\ell=1$ mode. Depending on the parameters, $\ell > 1$ modes can also have sufficient coherent shifts to drive instabilities with multiple coupled-bunch modes. This may help understanding the behavior of many coupled-bunch modes excited during the \gls{ptbl} instability~\cite{Gamelin2022}.

For simplicity, we will use the approximate case of a dipole instability in an equivalent quadratic potential with the same bunch length in a double-rf system at flat-potential, such as presented in Sec.~\ref{subsec:dipole-instability}. For this case, the coherent shift is $\Omega_{\mathrm{linear}, \ell}^2 = \langle\omega_s\rangle^2 + 2\langle\omega_s\rangle\Lambda_\ell^{(1)}$ with $\langle\omega_s\rangle = \alpha \sigma_\delta / \sigma_\tau$. The condition for the mode $\ell=1$ instability threshold condition will be set as $\Omega_{\mathrm{linear}, \ell=1}^2 \approx 0$, implying $\Re\left[\Lambda_{\ell=1}^{(1)}(0)\right] \approx -\langle\omega_s\rangle/2$. With that, we get the approximate threshold formula:
\begin{equation}
    \left[I_0\left(\frac{R}{Q}\right) \right]_\mathrm{threshold} \approx \frac{T_0 \sigma_\delta}{n^2 \sigma_\tau} \sqrt{\frac{E_0 \alpha \hat{V}_\rf}{2\pi eF_n h^3}}.\label{eq:threshold-formula}
\end{equation}
A mode-1 instability is expected for $I_0 (R/Q)$ values above this threshold. A similar formula was derived by Venturini (see slide 17 in Ref.~\cite{Venturini2020}). Interestingly, both formulas exhibit the scaling $\frac{\sigma_\delta}{n^2} \sqrt{E_0 \alpha \hat{V}_\rf / F_n}$. Venturini's formula is based on a dispersion-relation in a quartic rf potential for the dipole instability of the $\ell=1$ mode, obtained by calculating the intersection of the effective impedance with the stability diagram boundary. Hence, some differences from our formula are expected. Another formula was derived by T. He (see Eq.~(24) in Ref.~\cite{He2022b}), which exhibits a different scaling: ${\hat{V}_\rf / n^2 F_n}$. In Ref.~\cite{He2022a}, the significant impact of $\alpha$ and $\sigma_\delta$ on the \gls{ptbl} threshold was demonstrated, while T. He's threshold formula lacks an explicit dependence on $\sigma_\delta \sqrt{E_0 \alpha}$.

We do not expect that the formula Eq.~\eqref{eq:threshold-formula} can provide accurate absolute threshold values due to its various approximations. Nevertheless, it serves as an interesting result for exploring the dependence on relevant parameters and may be useful for comparing relative thresholds.
\bibliography{references.bib}
\end{document}